\documentclass[a4paper]{article}
\usepackage{listings}
\usepackage{xcolor}
\usepackage{multirow}
\usepackage{jheppub}
\usepackage{dcolumn}
\usepackage[english]{babel}
\usepackage[utf8x]{inputenc}
\usepackage[T1]{fontenc}
\usepackage{palatino}
\usepackage{amsmath}
\usepackage{afterpage}
\usepackage{graphicx, subcaption}
\usepackage[colorinlistoftodos]{todonotes}
\usepackage[colorlinks=true]{hyperref}
\usepackage{makeidx}
\newcommand{\be}{\begin{equation}}  
\newcommand{\ee}{\end{equation}}  
\newcommand{\bea}{\begin{eqnarray}}  
\newcommand{\eea}{\end{eqnarray}}  
\makeindex
\begin{document}

\vspace*{1.2cm}

\thispagestyle{empty}
\begin{center}
{\LARGE \bf Odderon observation: explanations and answers to questions/objections regarding the PRL publication}

\par\vspace*{7mm}\par

{

\bigskip

\large \bf Kenneth \"Osterberg on behalf of the D0 and TOTEM collaborations}

\bigskip

{\large \bf  E-Mail: kenneth.osterberg@helsinki.fi}

\bigskip

{Department of Physics and Helsinki Institute of Physics, University of Helsinki, Helsinki, Finland}

\bigskip

{\it Presented at the Low-$x$ Workshop, Elba Island, Italy, September 27--October 1 2021}

\vspace*{15mm}

\end{center}
\vspace*{1mm}

\begin{abstract}

The odderon observation recently published by the D0 and TOTEM collaborations has been widely accepted by a majority of the particle physics community and its importance recognized through dedicated physics seminars in the world major labs and physics institute. Naturally also some questions and objections have been raised, either privately or publicly, in discussion sessions and articles. In this proceedings article, a comprehensive list of these questions and objections are answered and supplementary material is provided. The methods and assumptions used in the extrapolation of the $pp$ elastic differential cross section to $\sqrt{s}$ = 1.96 TeV and its comparison to the D0 measurement in $p\bar{p}$ are shown to be valid and reasonable. Likewise, the methods and choices used for the $\rho$ measurement at LHC. Furthermore, objections against the odderon interpretation are demonstrated not to be valid. Finally, the combination of the different odderon significances, leading to the first experimental observation of odderon exchange, is shown to be well founded. 
\end{abstract}
 
 \section{Introduction}
 The D0 and TOTEM collaborations have recently published the observation of the odderon~\cite{Odderon-discovery}. The observation is based on combining two evidences for the odderon in complementary $|t|$-ranges using complementary TOTEM data sets: (1) a comparison of the proton-proton ($pp$) and proton-antiproton ($p\bar{p}$) elastic differential  cross sections ($d\sigma_{el}/dt$) in the $|t|$-range of the diffractive minumum ("dip") and the secondary maximum ("bump") of the $pp$ $d\sigma_{el}/dt$ at $\sqrt{s}$ = 1.96 TeV~\cite{Odderon-discovery} and (2) the total cross section ($\sigma_{tot}$) and $\rho$ measurements at \mbox{very low $|t|$ in $pp$ collisions at the LHC~\cite{TOTEM-rho-13TeV}.} The methods, assumptions and choices used in the analyses have raised questions that are answered in detail and supplementary material is provided in this proceedings contribution. Furthermore, the objections raised to the odderon interpretation of the evidences are shown not to be valid.

The explanations and answers are organized as follows. First the comparison of the $pp$ and $p\bar{p}$ $d\sigma_{el}/dt$ is briefly presented, then explanations regarding the $pp$ and $p\bar{p}$ comparison are provided and questions and objections raised are answered. Next the odderon evidence from the TOTEM $\rho$ and $\sigma_{tot}$ measurements is introduced and afterwards replies to the questions and objections raised regarding the analysis and interpretation are given. Finally the combination of the odderon signatures is discussed and responses to issues raised concerning the combination are provided.

\section{The comparison of elastic $pp$ and $p\bar{p}$ cross sections}
\begin{figure}
\begin{minipage}{0.55\linewidth}
\centerline{\includegraphics[width=0.925\linewidth]{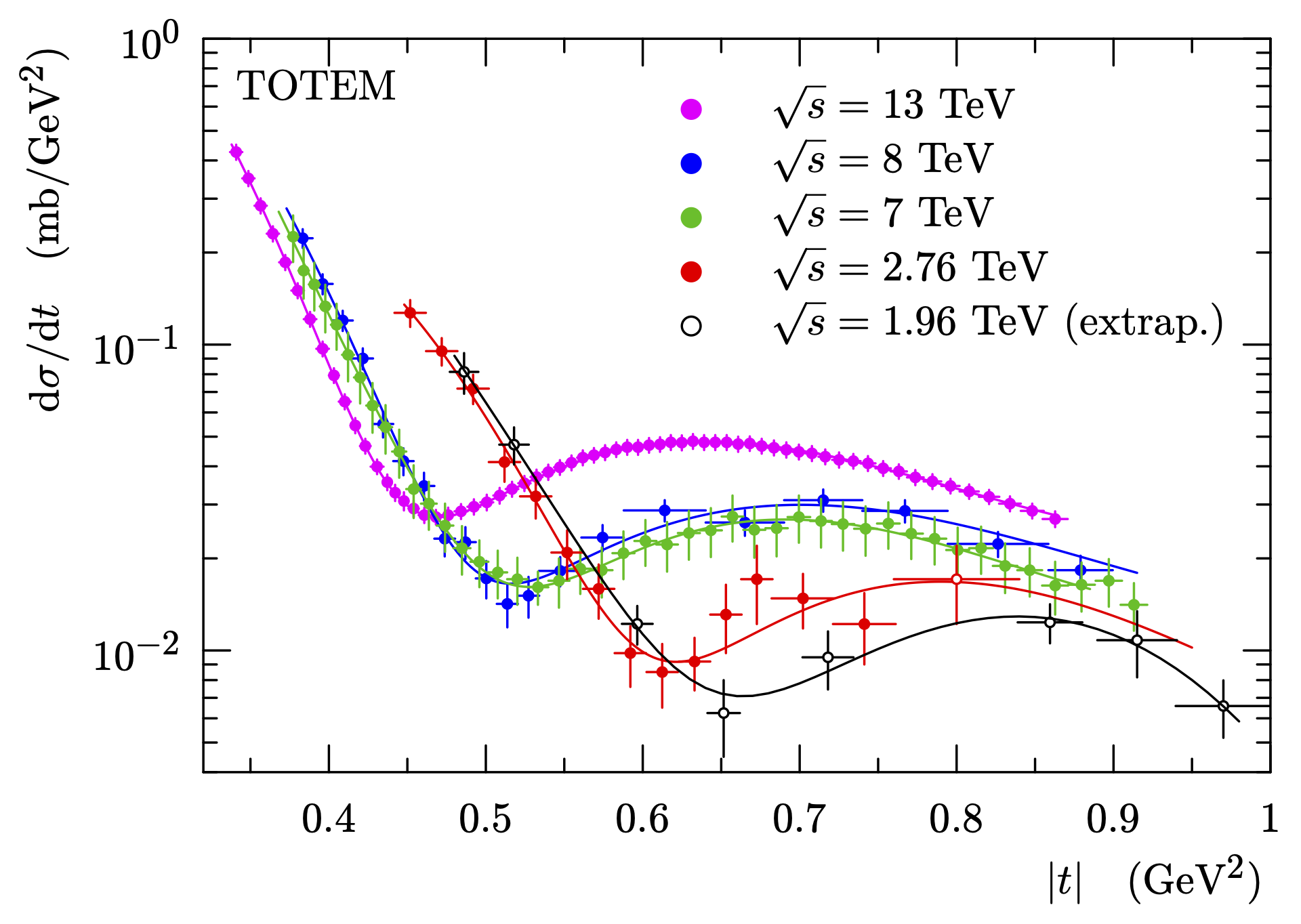}}
\end{minipage}
\hfill
\begin{minipage}{0.45\linewidth}
\centerline{\includegraphics[width=0.925\linewidth]{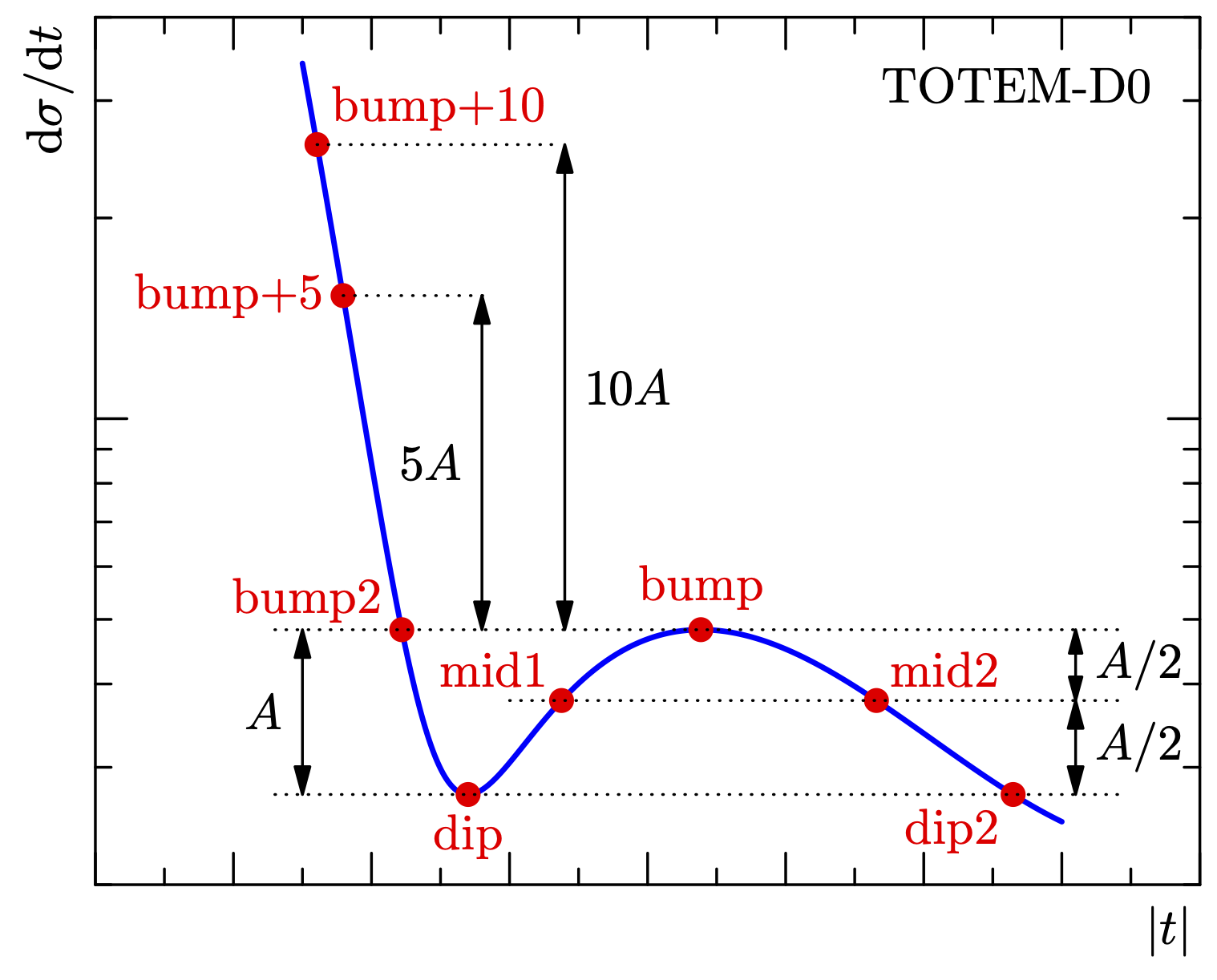}}
\end{minipage}
\hfill
\caption[]{Left: The TOTEM $pp$ elastic cross sections at 2.76, 7, 8, and 13 TeV (full circles), and
the extrapolation to 1.96 TeV (empty circles). Right: Schematic definition of the characteristic points in the TOTEM differential cross section data. $A$ represents the vertical bump to dip distance.}
\label{fig:dsigmadt}
\end{figure}

Each $pp$ $d\sigma_{el}/dt$ measurement at TeV energy scale shows a characteristic dip, followed by a bump, as illustrated by Fig.~\ref{fig:dsigmadt} (left), whereas the $p\bar{p}$ $d\sigma_{el}/dt$ at TeV energy scale only exhibits a flat behaviour in the region of the expected positions of the dip and bump. This difference in the $pp$ and $p\bar{p}$ $d\sigma_{el}/dt$ would naturally occur for $t$-channel odderon exchange, since at the dip the dominant pomeron exchange is largely suppressed, and the odderon amplitude can play a significant role. Contrary to the pomeron amplitude, the odderon amplitude has a different sign for $pp$ and $p\bar{p}$. 

\begin{figure}
\begin{center}
\centerline{\includegraphics[width=0.60\linewidth]{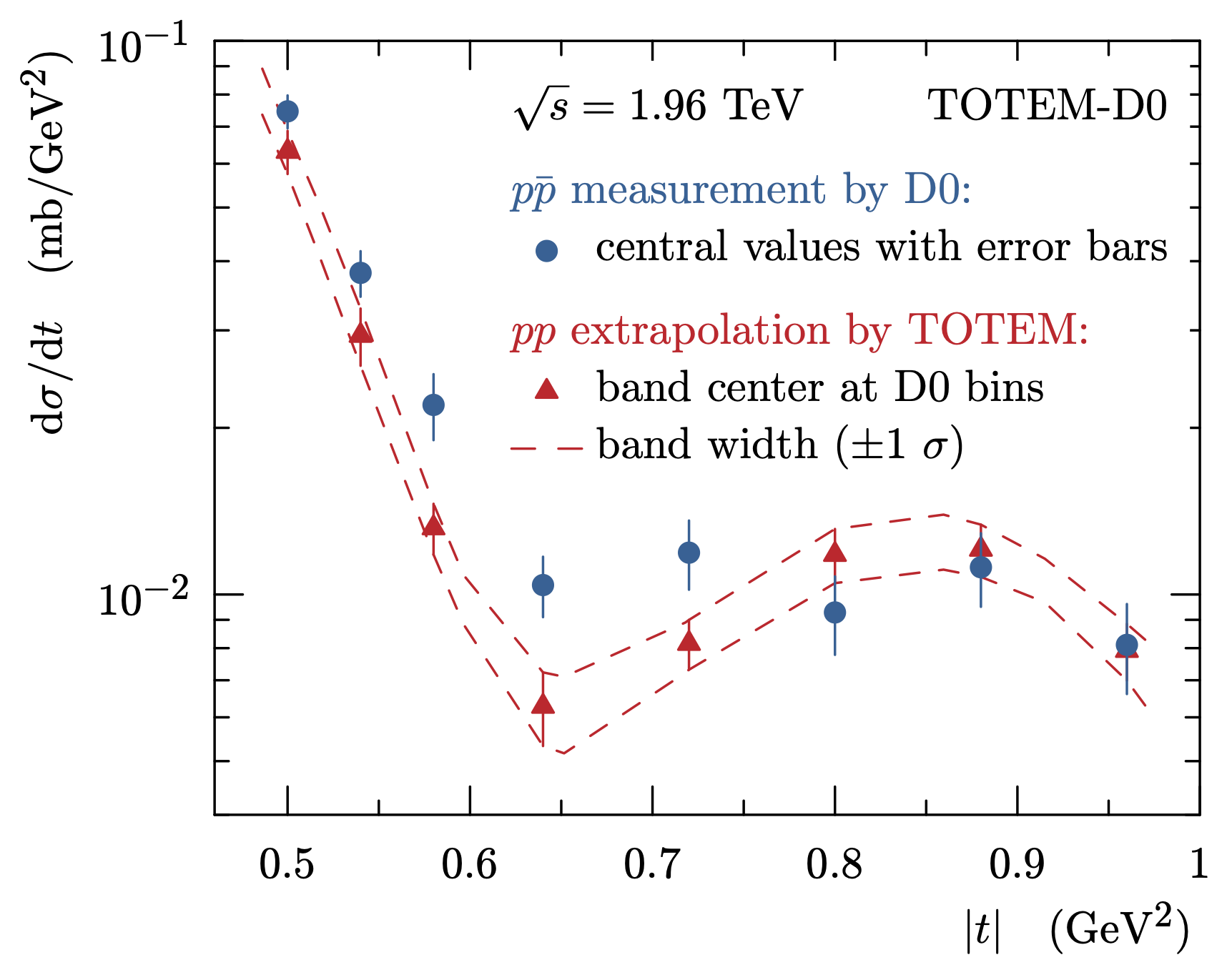}}
\caption{Comparison between the D0 $p\bar{p}$ measurement at 1.96 TeV and the extrapolated TOTEM $pp$ cross section (with its 1$\sigma$ uncertainty band), rescaled to match the D0 optical point. Note that the uncertainties at different $|t|$ values in the 1$\sigma$ uncertainty band are strongly correlated.}
\label{fig:dsigmadt_comp}
\end{center}
\end{figure}

To quantify the difference, eight characteristic points in the region of the dip and the bump, shown in Fig.~\ref{fig:dsigmadt} (right), of the TOTEM 2.76, 7, 8, and 13 TeV $pp$  $d\sigma_{el}/dt$ are extrapolated using a data-driven approach to obtain the 1.96 TeV $pp$ $d\sigma_{el}/dt$. The observed difference of 3.4$\sigma$ significance between the extrapolated $pp$ and the D0 $p\bar{p}$ $d\sigma_{el}/dt$ at 1.96 TeV in the region of the dip and the bump of the $pp$  $d\sigma_{el}/dt$, as shown by Fig.~\ref{fig:dsigmadt_comp}, is interpreted as evidence for odderon exchange. Note that the comparison is made in a common $t$-range (0.50 $\leq |t| \leq$ 0.96 GeV$^2$) of the $pp$ and $p\bar{p}$ $d\sigma_{el}/dt$.

\subsection{Questions and objections raised regarding the analysis and the interpretation}

\begin{figure}
\begin{center}
\centerline{\includegraphics[width=0.85\linewidth]{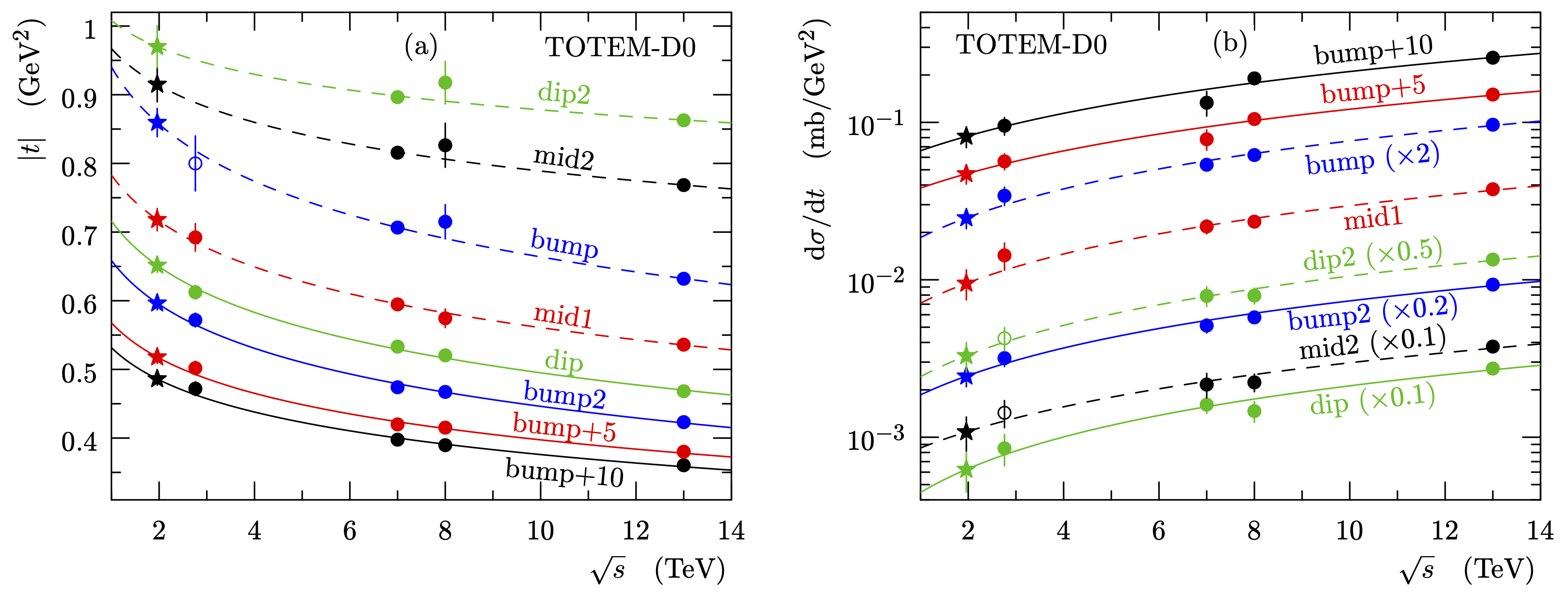}}
\caption{Characteristic points in (a) |t| and (b) $d\sigma / dt$ from TOTEM measurements at 2.76, 7, 8 and 13 TeV (circles) as functions of $\sqrt{s}$ extrapolated to 1.96 TeV (stars). Filled symbols are from measured pints; open symbols are from extrapolations or definitions of the characteristic points.}
\label{fig:dsigmadt_extra}
\end{center}
\end{figure}

A first objection that has been raised is a possible model dependence introduced by the formulas, $|t| = a \log (\sqrt{s} {\rm [TeV]}) + b$ and $d\sigma /dt = c \sqrt{s} {\rm [TeV]} + d$, used to extrapolate the TOTEM measured $|t|$  and differential cross section ($d\sigma /dt$)  values at 2.76, 7, 8 and 13 TeV to 1.96 TeV to obtain the characteristic points of the $pp$ $d\sigma /dt$ at 1.96 TeV, see Fig.~\ref{fig:dsigmadt_extra}. Firstly, it should be noted that the $\sqrt{s}$ range of the extrapolation from 2.76 TeV is small, only about 8 \%, compared to the $\sqrt{s}$ range that the validity of formulas are tested with the fits. Secondly, for each characteristic point, the closest measured point to the characteristic point in terms of $d\sigma /dt$ is used as measured and if two adjacent points have about equal $d\sigma /dt$, the two bins are merged avoiding any model-dependent extrapolation between bins. Thirdly, having 3-4 data points limits the extrapolation formulas to ones with maximally two parameters. Alternative functional forms with other $\log \sqrt{s}$ or $\sqrt{s}$ powers yield extrapolated values at 1.96 TeV well within the uncertainties of the extrapolated values given by the fits using the above $\sqrt{s}$ dependence for $|t|$ and $d\sigma /dt$. Fourthly, it is not obvious that the same functional form would give good fits for all characteristics points both in $|t|$ and $d\sigma /dt$ (majority of $\chi^2$ values $\sim$1 per degree of freedom (d.o.f.)) that probably is related to some general energy independent properties of elastic scattering, see e.g. Refs.~\cite{Martynov_Nicolescu, Durham}. So if there is any model dependence at all, it is largely contained in the quoted uncertainties, in particular due to short extrapolation range and the generality of the functional form used for extrapolating the characteristic points. Note also that the shape and hierarchy of the extrapolated $pp$ $d\sigma /dt$ w.r.t. the measured $pp$ $d\sigma /dt$ is preserved as shown by Fig.~\ref{fig:dsigmadt} (left), i.e. a constant bump-to-dip $d\sigma /dt$ ratio with energy, a descreasing $|t|$ of the diffractive cone, dip and bump position with energy and decreasing values of the $d\sigma /dt$'s in the dip-bump region with energy. Extrapolating the measured cross sections is more robust that fitting the $pp$ $d\sigma /dt$ at each energy and extrapolating the fit parameters, which tend to compensate each other and whose correlations might be different at different energies. 

\begin{figure}
\begin{center}
\centerline{\includegraphics[width=0.55\linewidth]{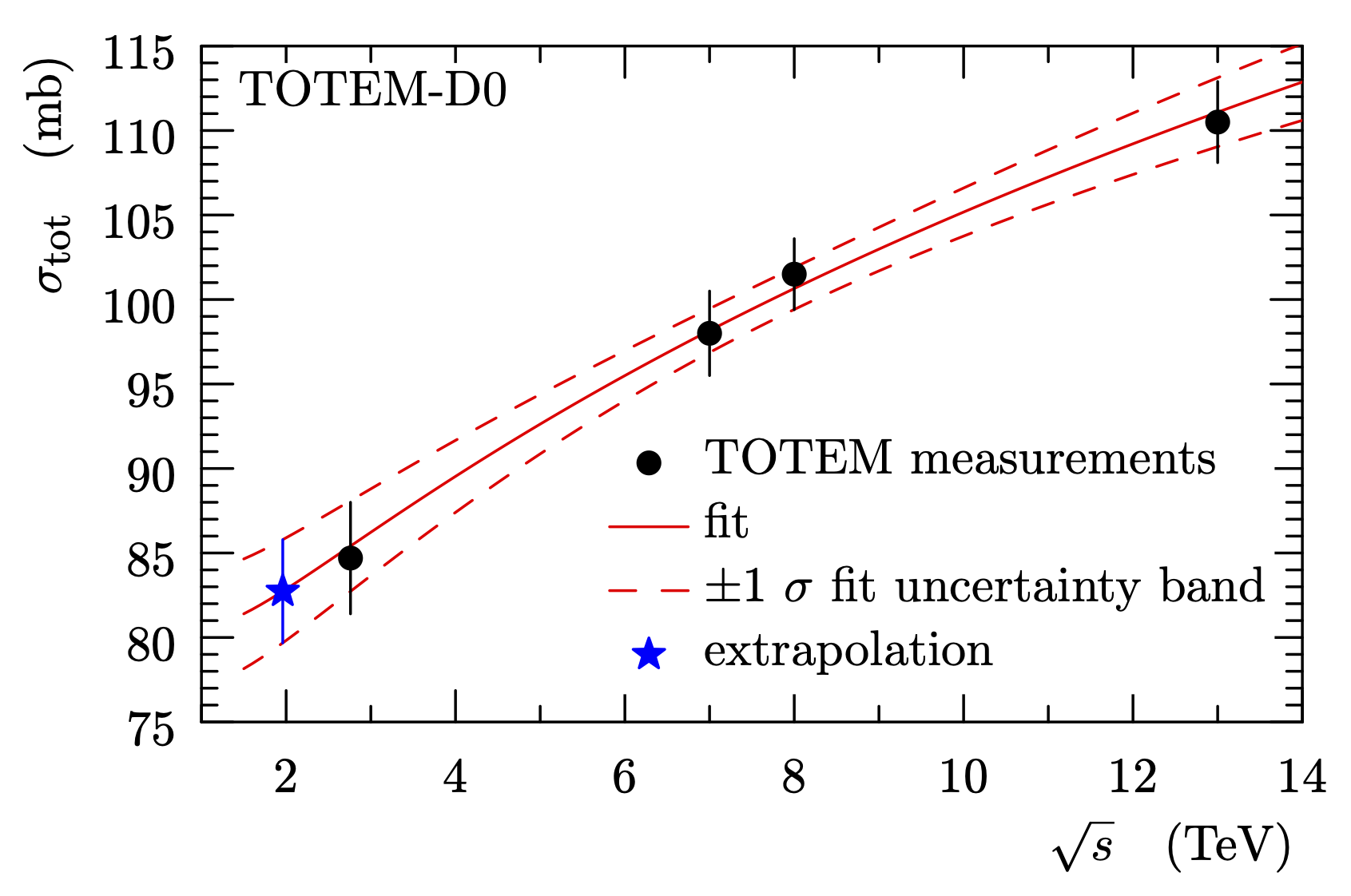}}
\caption{The $\sigma_{tot}$ from TOTEM measurements at 2.76, 7, 8 and 13 TeV (circles) as a functions of $\sqrt{s}$ extrapolated to the center-of-mass energy of the D0 measurement (star).}
\label{fig:sigmatot_extra}
\end{center}
\end{figure}

A similar objection has been raised concerning a possible model dependence introduced by the formula, $\sigma_{tot} = b_1 \log^2 (\sqrt{s} {\rm [TeV]}) + b_2$, used to extrapolate the TOTEM measured total cross section ($\sigma_{tot}$)  values at 2.76, 7, 8 and 13 TeV to 1.96 TeV as shown Fig.~\ref{fig:sigmatot_extra}, obtaining $\sigma_{tot} (pp)$ = 82.7 $\pm$ 3.7 mb at 1.96 TeV. Here the argumentation is similar to the one for the $|t|$ and $d\sigma/dt$ values. Firstly, the $\sqrt{s}$ range of the extrapolation is small, only about 8 \%, compared to the $\sqrt{s}$ range that the validity of formula is tested with the fit. Secondly, having four data points limits the extrapolation formulas to ones with maximally three parameters. Alternative functional forms such as $\log^2\sqrt{s} + \log\sqrt{s} + C$, $s + \sqrt{s} + C$ or  $s^{1/4} + C$ gave extrapolated values at 1.96 TeV well within the quoted $\sigma_{tot} (pp)$ uncertainty. Thirdly, the fit to the TOTEM $\sigma_{tot}$ measurements gives a $\chi^2$ per d.o.f. smaller than 1. So in conclusion, if there is any model dependence, it is well within the quoted uncertainty. Note that 1.96 TeV is in a boundary region for $\sigma_{tot}$, dominated by a $\log \sqrt{s}$ dependence for lower energies and a $\log^2 \sqrt{s}$ dependence for higher energies. Therefore the extrapolation of the TOTEM $\sigma_{tot}$ measurements is only valid for $\sqrt{s} \ge $ 1 TeV, which is sufficient for the purpose above. 

Also a somewhat similar objection has been raised concerning the interpolation of the characteristic points of the $pp$ $d\sigma_{el} /dt$ at 1.96 TeV to the $|t|$ values of the measured D0 $p\bar{p}$ $d\sigma_{el} /dt$ in the range 0.50 $\le |t| \le$ 0.96 GeV$^2$ using the double exponential:
\begin{eqnarray}
h (t) & = &  a_1 e^{-a_2 |t|^2 - a_3 |t|} +  a_4 e^{-a_5 |t|^3 -a_6 |t|^2 - a_7 |t|} \, \,  , 
\label{eq:double_exp}
\end{eqnarray}
where the first exponential describes the diffractive cone (with a steeper slope towards the dip) and the second exponential the asymmetric bump structure and subsequent falloff. The fit to the characteristic points of the $pp$ $d\sigma_{el} /dt$ at 1.96 TeV using Eq.~\ref{eq:double_exp},  gives a $\chi^2$ per d.o.f. smaller than 1. The same functional form describes well the measured $pp$ $d\sigma_{el} /dt$ in the dip and bump region for at 2.76, 7, 8 and 13 TeV, as illustrated in Fig.~\ref{fig:dsigmadt} (left), with a $a_4$ term i.e. bump term significantly different from zero. This reassures that Eq.~\ref{eq:double_exp} can be safely used for the interpolation given that the functional form corresponds to a very distinct shape of the $d\sigma_{el} /dt$. The interpolation uncertainty is evaluated using a MC simulation where the cross section values of the eight uncorrelated characteristic points at 1.96 TeV are varied within their Gaussian uncertainties and new fits given by Eq.~\ref{eq:double_exp} are performed. This provides a $pp$ cross section value at each $|t|$ value that was checked to correspond to a Gaussian distribution with the quoted uncertainty. All of this suggests that the model dependence due to the interpolation must be well within the quoted uncertainty.

Another objection is the assumption that the optical points (OP) (${d\sigma_{el}/dt}\left \vert_{t = 0} \right.$) of $pp$ and $p\bar{p}$ are equal. The basis is the Pomeranchuk theorem~\cite{Pomeranchuk} stating that the ratio of the $pp$ and $p\bar{p}$ $\sigma_{tot}$ is 1, when $\sqrt{s}$ approaches infinity. Using the optical theorem, this leads to the ratio of the OPs of  $pp$ and $p\bar{p}$ to be 1, when $\sqrt{s}$ approaches infinity. This doesn't imply that they are necessarily equal, however any possible difference between them must be due to the C-odd amplitude, which in the TeV-range is due to the odderon, since secondary reggeons can safely be ignored due to the decrease of their amplitude with $\sqrt{s}$, whereas the odderon amplitude is expected to increase with $\sqrt{s}$~\cite{Reggeons}. Therefore the assumption of equal $pp$ and $p\bar{p}$ OP is valid as long as the maximal possible odderon effect on the $\sigma_{tot}$ and hence on the OP is included as a systematic uncertainty for the OP. 

The assumption of equal $pp$ and $p\bar{p}$ OP can be tested comparing the extrapolated ${d\sigma_{el}/dt}\left \vert_{t = 0} \right.$ = 357 $\pm$ 26 mb/GeV$^2$ at 1.96 TeV with the extrapolation of the D0 $d\sigma^{p\bar{p}}_{el}/dt$ measurement to $|t| = 0$ obtaining  ${d\sigma_{el}/dt}\left \vert_{t = 0} \right.$ = 341 $\pm$ 49 mb/GeV$^2$. As can be noted they numerically agree well within the uncertainties, in fact the $p\bar{p}$ OP and its uncertainty encompasses the $pp$ OP and its uncertainty. 

Since the $pp$ and $p\bar{p}$ OP measurements measure the same physics quantity in the assumption of equal $pp$ and $p\bar{p}$ OP, one can estimate a weighted average from them and conclude that the precision on the common OP is determined by the measurement with the better precision, i.e. the $pp$ OP. Therefore the uncertainty on the $p\bar{p}$ OP can be ignored, since the uncertainty of two independent measurement of the same quantity never can be larger than the smaller of the two uncertainties. This procedure is still valid even if the $pp$ and $p\bar{p}$ OP would correspond to two different physics quantities with a known difference as long as the difference is included in the overall uncertainty. The maximal possible difference due to odderon exchange on the OP is estimated from the maximal odderon model to be 2.9 \% at 1.96 TeV that is added in quadrature to the uncertainty of the experimental $pp$ OP to give an overall 7.4 \% relative uncertainty on the common OP. Effects on the OP from secondary reggeons and from differences between the $pp$ and $p\bar{p}$ $\rho$ values are negligible.

Also the ability to extrapolate the D0 $d\sigma^{p\bar{p}}_{el}/dt$ to the OP has been questioned, since the measurement only covers $|t|$-values down to 0.26 GeV$^2$. In particular, since the $B$-slope measurements in $p\bar{p}$ at 0.546 TeV seems to indicate that the $B$-slope is 10-15 \% steeper for low $|t|$-values ($\lesssim$ 0.15 GeV$^2$) than higher $|t|$-values~\cite{UA4_slope}. However neither CDF~\cite{CDF_slope} nor E710~\cite{E710_slope1} observe any indication of a change of $B$-slope of that size below $|t|$ = 0.25 GeV$^2$ at 1.8 TeV.  
Even if the difference between the central values of the two E710 $B$-slope measurements~\cite{E710_slope1, E710_slope2} would be interpreted as an actual $B$-slope difference as a function of $|t|$, the change on the OP would be much smaller ($\sim$ 4 \%) than the luminosity uncertainty of 14.4 \% that dominates the D0 $p\bar{p}$ OP. Comparing TOTEM $\sigma_{tot}$ measurements at $\sqrt{s}$ = 8 and 13 TeV in $pp$ based on $B$-slopes extracted from data with and without acceptance in the Coulomb Nuclear Interference (CNI)-region, the ones with CNI-region data give about 1 \% higher $\sigma_{tot}$ thus about 2 \% higher OP (and steeper $B$-slope). So there is no indication that the D0 $p\bar{p}$ OP cannot be trusted. Note that if the pattern from 1.8 TeV $p\bar{p}$ and 8 and 13 TeV $pp$ measurements at low $|t|$ would be used to correct the D0 $p\bar{p}$ OP at $\sqrt{s}$ = 1.96 TeV for possible bias due to lack of such data, the central values of the D0 $p\bar{p}$ and TOTEM $pp$ OP's would be even closer.   

As a result of the interpolation from the characteristic points of the extrapolated $pp$ $d\sigma /dt$ to the $|t|$ values of the D0 $p\bar{p}$ $d\sigma /dt$, the $pp$ $d\sigma /dt$ at the $|t|$ values of the $p\bar{p}$ $d\sigma /dt$ are strongly correlated implying that the full covariance matrix of the $pp$ data points must be included in the $\chi^2$ for the comparison of the $pp$ and $p\bar{p}$ $d\sigma /dt$. The $\chi^2$-formula used:

\begin{eqnarray}
\chi^2 & = &  \sum_{i,j=1}^8 \left\{ \left( \frac{d\sigma^{pp, norm}_{el,i}}{dt} - \frac{d\sigma^{p\bar{p}}_{el,i}}{dt} \right) C^{-1}_{i,j} \left( \frac{d\sigma^{pp, norm}_{el,j}}{dt} - \frac{d\sigma^{p\bar{p}}_{el,j}}{dt} \right) \right\} + \frac{(A-A_0)^2}{\sigma_A^2} + \frac{(B-B_0)^2}{\sigma_B^2} \, \,  , 
\label{eq:chi2}
\end{eqnarray}

where $C_{i,j}$ is the covariance matrix, $A$ and $B$ are the two constraints and $d\sigma^{pp, norm}_{el,i} /dt$ is the $pp$ $d\sigma_{el}/dt$ normalized to the $p\bar{p}$ integral elastic cross section ($\sigma_{el}$) in the $|t|$ range of the comparison. The first constraint ($A$) is the normalization due to the matching of the $pp$ and $p\bar{p}$ OPs. The second constraint ($B$) is the matching of the $pp$ and $p\bar{p}$ $B$-slopes in the diffractive cone. The Pomeranchuk and the optical theorem infer that the ratio of the $pp$ and $p\bar{p}$ total $\sigma_{el}$ should be 1, when $\sqrt{s}$ goes to infinity. From this, one can deduce that the ratio of the $pp$ and $p\bar{p}$ elastic $B$-slopes should be 1, when $\sqrt{s}$ approaches infinity, since the $\sigma_{el}$ in the Coulomb region and in the region beyond the dip is negligible compared the one in the diffractive cone and the $d\sigma_{el}/dt$ in the diffractive cone is described by $e^{-B|t|}$~\cite{Cornille_Martin}. This doesn't imply that they are exactly equal but any difference between the $pp$ and $p\bar{p}$ elastic $B$-slopes at the TeV-scale is due to the odderon. Since the pomeron dominates in the diffractive cone region at 1.96 TeV, the $B$-slopes of $pp$ and $p\bar{p}$ are expected to be equal. This is verified to be true within the experimental uncertainties for the D0 $p\bar{p}$ and the TOTEM $pp$ $B$-slopes.

Therefore Eq.~\ref{eq:chi2} expresses the complete $\chi^2$, including the covariance matrix and the terms for the fully correlated uncertainties, thus also expressing the exact number of d.o.f. Eq.~\ref{eq:chi2} gives for six d.o.f. a significance of 3.4$\sigma$ for the difference between the TOTEM $pp$ and the D0 $p\bar{p}$ $d\sigma_{el}/dt$ at 1.96 TeV using the eight points in the region of the dip and the bump. The $\chi^2$ and therefore the significance is largely dominated by the first term of Eq.~\ref{eq:chi2} related to the shape of the $d\sigma_{el}/dt$. The obtained significance is confirmed by a Kolmogorov-Smirnov test of the difference between the $pp$ and $p\bar{p}$ $d\sigma_{el}/dt$ in the same $|t|$ range, where the correlations of the data points are included using Cholesky decomposition~\cite{Cholesky} and the normalisation difference via Stouffer's method~\cite{Stouffer}. 

\section{The TOTEM $\rho$ and $\sigma_{tot}$ measurements}

The second evidence of odderon exchange in elastic scattering is from the measurements of $\rho$, the ratio of the real and imaginary part of the elastic hadronic amplitude at $t$ = 0, and $\sigma_{tot}$ in $pp$ collisions at the LHC~\cite{TOTEM-rho-13TeV}. Models~\cite{Compete, Durham, Block_Halzen} are unable describe both the TOTEM $\sigma_{tot}$ and $\rho$ measurements without including odderon exchange. The disagreement between the measurements and the models is between 3.4$\sigma$ and 4.6$\sigma$ depending on the model. Comparison between the predictions of the COMPETE models~\cite{Compete} and the TOTEM $\sigma_{tot}$ and $\rho$ measurements is shown in Fig.~\ref{fig:compete_sigmatot_rho}. Note that the COMPETE~\cite{Compete} and Block-Halzen~\cite{Block_Halzen} models include secondary Reggeon-like C-odd terms proportional to $\sim 1/\sqrt{s}$ to describe the difference of $pp$ and $p\bar{p}$ scattering below 0.1 TeV that should not be confused with odderon-like C-odd terms that are expected to increase with $\sqrt{s}$.

When comparing different $\rho$ measurements, it is important to make sure that the prescriptions (functional form for the hadronic amplitude and the phase, CNI formula and $|t|$-range) used in the extraction are as similar as possible, otherwise it doesn't necessarily lead to the same physics quantity. This is especially true in the comparison with previous $\rho$ measurements. The $\sqrt{s}$ trend in the TeV range predicted by odderon exchange~\cite{Martynov_Nicolescu, Durham13TeV} is observed for the most precise $\rho$ measurements for $pp$ and $p\bar{p}$ in the TeV range, when extracted using the same prescription:  $\rho$ = 0.135 $\pm$ 0.015 at 0.546 TeV in $p\bar{p}$~\cite{UA4_rho} and $\rho$ = 0.09 $\pm$ 0.01 at 13 TeV in $pp$~\cite{TOTEM-rho-13TeV}. Note also that several groups, including A. Donnachie and P.V. Landshoff~\cite{Donnachie_Landshoff} and J.R. Cudell and O.V. Selyugin~\cite{Cudell_Selyugin}, have obtained compatible $\rho$ values (in the range 0.08-0.10), when taking the TOTEM 13 TeV CNI data as given and using a similar prescription as TOTEM~\cite{TOTEM-rho-13TeV}, contrary to the results they quote when they misinterpret or allow themselves the freedom to shift the TOTEM data and related uncertainties.

\subsection{Questions and objections raised concerning the analysis and the interpretation}

The authors of the PDG review of High Energy Soft QCD and Diffraction~\cite{PDG} claim that analyzing the whole ensemble of TeV-range elastic $pp$ and $p\bar{p}$ low $|t|$ data including the TOTEM measurements at LHC, a reasonable description can be obtained using a C-even amplitude (pomeron) only, that is, without an odderon, in contradiction with the conclusion by TOTEM. This statement does not hold once one start to examine the exact predictions. For example, the model of the authors~\cite{Durham} fails to describe both the TOTEM $\rho$ and $\sigma_{tot}$ measurements in $pp$ at 7, 8 and 13 TeV ($\sim$ 3.4$\sigma$ difference) and especially the elastic $d\sigma/dt$ in $p\bar{p}$ for the dip and bump region at 1.96 TeV ($\sim$ 4.3$\sigma$ difference). A good description of the LHC $pp$ data without the odderon, leads inevitably to a significantly worse description of the Tevatron $p\bar{p}$ data and vice versa.

\begin{figure}
\begin{center}
\centerline{\includegraphics[width=0.735\linewidth]{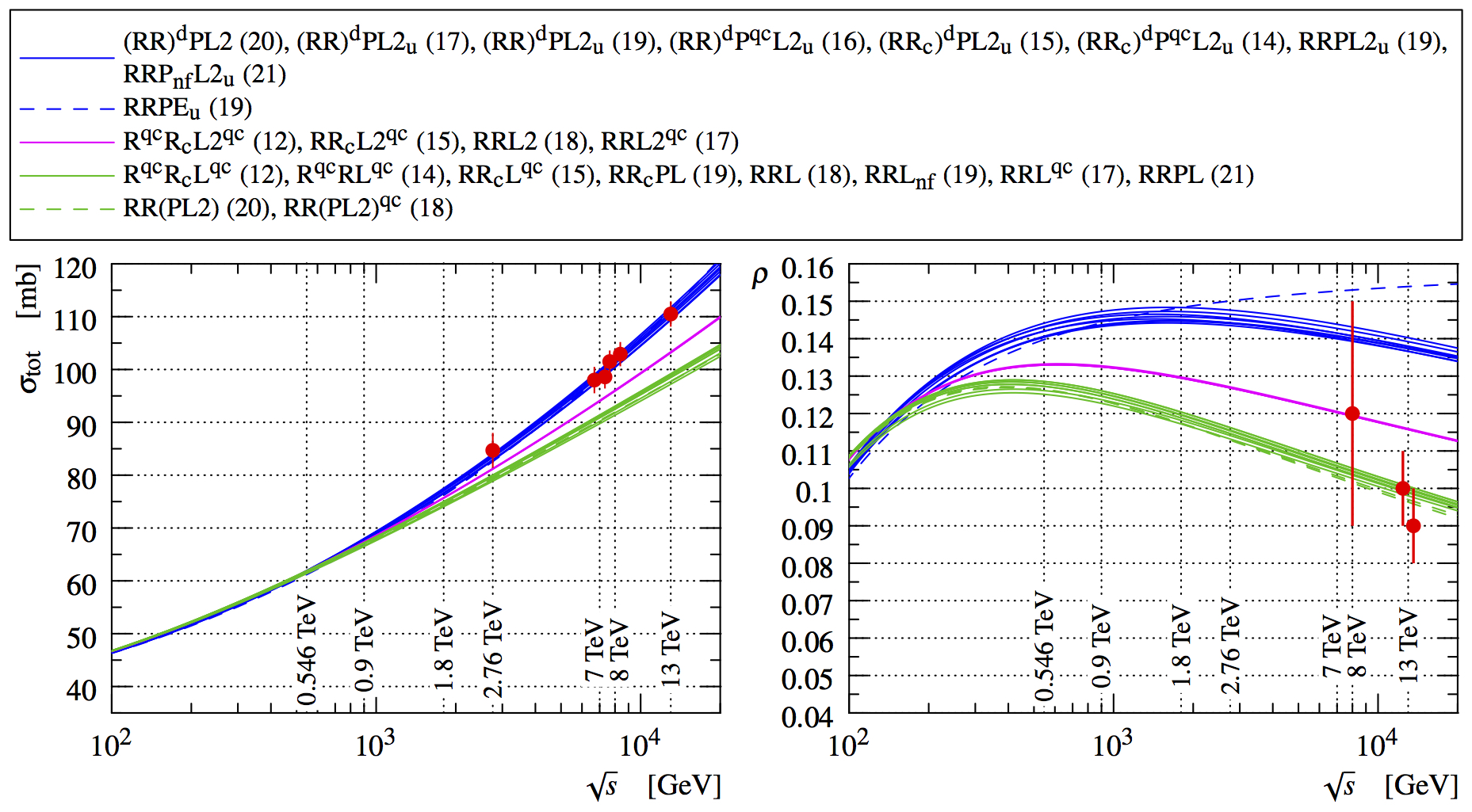}}
\caption{Predictions of the $pp$ total cross section ($\sigma_{tot}$) and $\rho$ parameter as function of $\sqrt{s}$ by each COMPETE~\cite{Compete} model (see legend for model) with the TOTEM measurements marked in red.}
\label{fig:compete_sigmatot_rho}
\end{center}
\end{figure}

\begin{figure}
\begin{minipage}{0.50\linewidth}
\centerline{\includegraphics[width=0.885\linewidth]{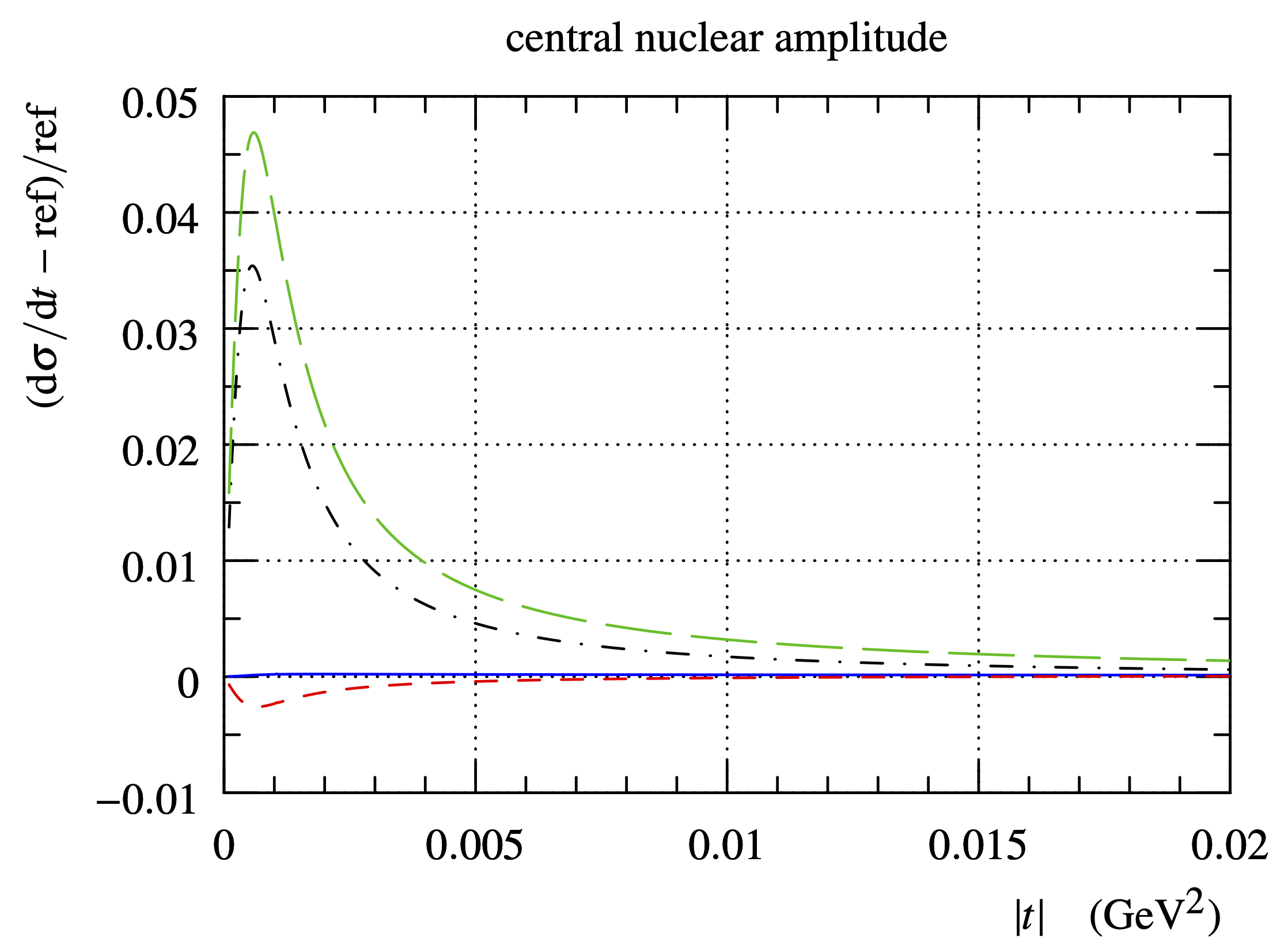}}
\end{minipage}
\hfill
\begin{minipage}{0.50\linewidth}
\centerline{\includegraphics[width=0.885\linewidth]{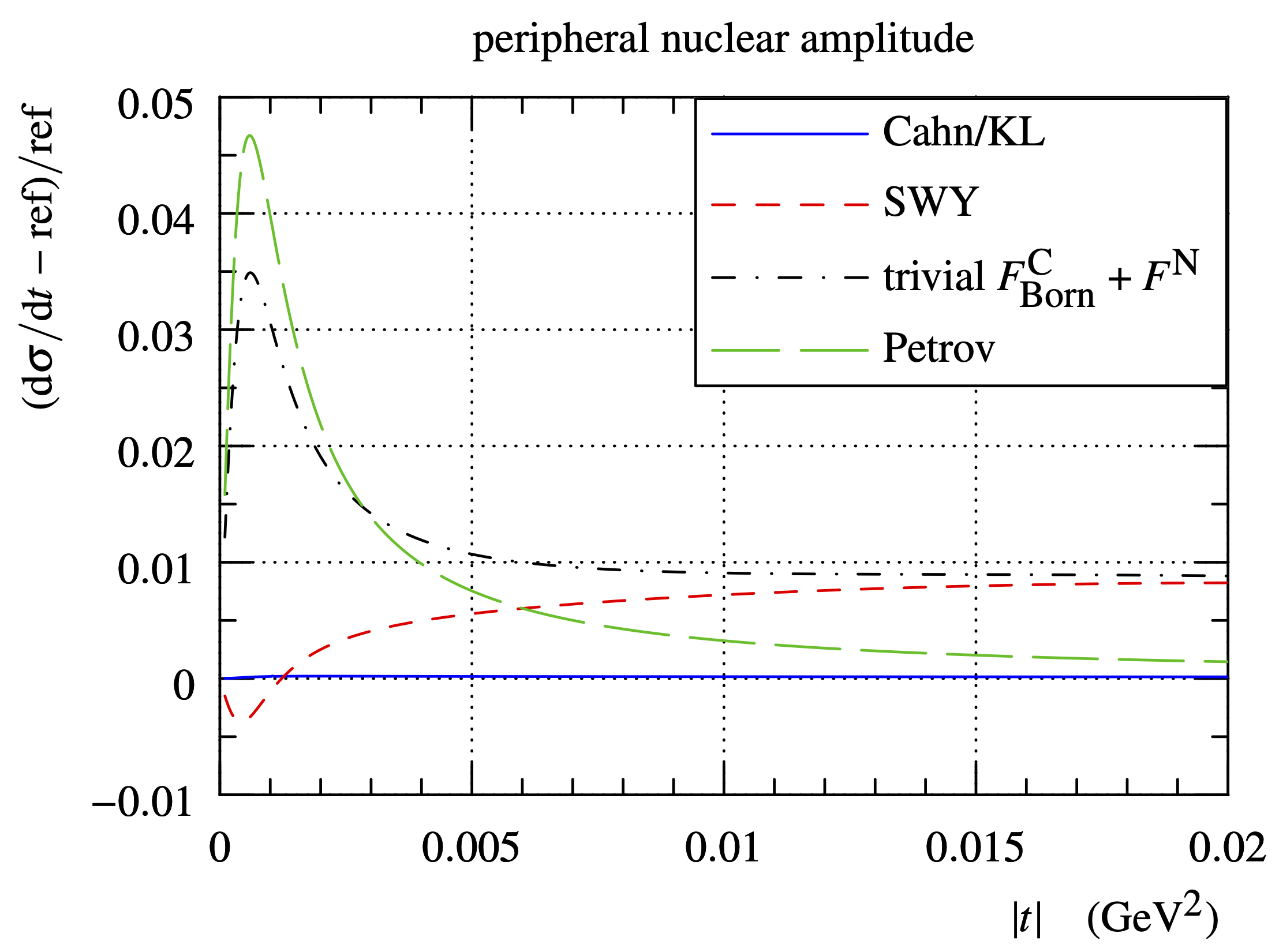}}
\end{minipage}
\hfill
\caption[]{The relative difference of the differential elastic cross section in the CNI region between various CNI formulae (see text) and the numerical calculation of the Coulomb and nuclear eikonals to all orders of $\alpha$ (denoted "ref") ~\cite{Kaspar} for central (left) and peripheral (right) nuclear amplitudes. The labelling refects the impact parameter behaviour: central nuclear amplitudes yield profile functions peaking at smaller impact parameter value than peripheral amplitudes.}
\label{fig:Kaspar_diff}
\end{figure}

In the PDG review, the statement of the authors is backed up by a similar attempt by Donnachie-Landshoff~\cite{Donnachie_Landshoff}, which claim to describe the elastic $d\sigma/dt$ data at small $|t|$ from 13.76 GeV to 13 TeV without the odderon. Donnachie-Landshoff obtain a $\rho$ = 0.14 in $pp$ at $\sqrt{s}$ = 13 TeV, when using the TOTEM 8 TeV CNI data~\cite{TOTEM-rho-8TeV} in addition to the TOTEM 13 TeV CNI data~\cite{TOTEM-rho-13TeV}, whereas when using only the TOTEM 13 TeV CNI data they obtain a $\rho$ = 0.10. This is not possible, if experimental uncertainties are treated correctly, since the TOTEM 13 TeV CNI data is about a factor three more precise than the TOTEM 8 TeV CNI data when the normalisation uncertainty is not taken into account. It is likely that the normalisation uncertainty that is common to all data points has not been treated correctly as a separate term $A$ in the $\chi^2$ as in Eq.~\ref{eq:chi2} in the fits by Donnachie-Landshoff. Otherwise one cannot explain the large weight the TOTEM 8 TeV CNI data obtains in the Donnachie-Landshoff fits. The normalisation uncertainty is the dominating uncertainty in the TOTEM CNI data except at the smallest $|t|$ values and  smaller in the TOTEM 8 TeV CNI data than in the TOTEM 13 TeV CNI data, 4.2 \% compared to 5.5 \%. Note also that in Ref.~\cite{Donnachie_Landshoff}, a trivial sum of Coulomb and nuclear elastic amplitudes is used, ignoring completely CNI effects on the amplitude, leading to relative deviations in the total elastic amplitude of several percent, see Fig.~\ref{fig:Kaspar_diff}.   

The PDG review also states that the model RR(PL2)$^{qc}$ of COMPETE (dashed green line in Fig.~\ref{fig:compete_sigmatot_rho}) is consistent with the TOTEM 13 TeV $\rho$ and $\sigma_{tot}$ within 1$\sigma$~\cite{Cudell_Selyugin}, in contradiction with the statement that all COMPETE models are incompatible with the TOTEM $\rho$ and $\sigma_{tot}$ measurements. This agreement with the RR(PL2)$^{qc}$ model is obtained by modifying the normalization of the TOTEM 13 TeV elastic $d\sigma/dt$ by $\sim$2$\sigma$ (when including the Coulomb normalization that was not taken into account in Ref.~\cite{Cudell_Selyugin}). Since the normalization of the TOTEM 13 TeV CNI data is obtained from two completely independent data sets and methods (optical theorem and Coulomb amplitude) that agree very well, it is unlikely that it is off by $\sim$2$\sigma$. The standard approach in physics is not to modify the data but instead adjust the model to describe the data and not vice versa. Without modifying the normalization of the TOTEM 13 CNI data, the original version of the RR(PL2)$^{qc}$ model~\cite{Compete} fails to describe the $\sigma_{tot}$ in pp at $\sqrt{s}$ = 2.76, 7, 8 and 13 TeV ($\sim$5.4$\sigma$ difference).

Regarding the determination of $\rho$, it important to stress that most of the sensitivity to $\rho$ is contained in only a few data bins in the CNI region, between those at very low $|t|$ with a significantly larger Coulomb than CNI contribution and the large majority of data bins at higher $|t|$, where the hadronic amplitude dominates. Experience from TOTEM has shown that the fits should be done in several steps in separate $|t|$ ranges, first to fix the other parameters (hadronic amplitude and Coulomb normalisation) before the $\rho$ to avoid any bias in the $\rho$ determination from data bins with very little or without any sensitivity to $\rho$, see e.g. section 6.3 in Ref.~\cite{TOTEM-rho-13TeV}. In Refs.~\cite{Donnachie_Landshoff, Cudell_Selyugin} it is not stated whether the fits have been performed in several steps to avoid bias in the $\rho$ determination from data bins with minimal sensitivity to $\rho$ or whether they have been performed in a single step.   

In addition, the CNI formulae of Cahn~\cite{Cahn} and Kundr{\' a}t-Locaji{\v c}ek (KL)~\cite{KL} used for the $\rho$ determination at 13 TeV have been claimed to contain flaws including inexact approximation of the Coulomb amplitude and too early truncation of the power series of the electromagnetic coupling $\alpha$~\cite{Petrov}. A numerical calculation of the Coulomb and nuclear eikonals to all orders of $\alpha$~\cite{Kaspar} verified that the CNI formulae of Cahn and KL reproduce the numerical estimate for the phase and the $d\sigma/dt$ at a precision significantly below the current experimental one, as shown by Fig.~\ref{fig:Kaspar_diff}. Hence any approximations done by Cahn and KL do not have any detrimental effect on the $\rho$ determination. Instead, the CNI formula of Ref.~\cite{Petrov} and the sum of Coulomb and nuclear amplitudes~\cite{Godizov} were found to deviate from the numerical estimate by several percent. The SWY formula~\cite{SWY} reproduces the numerical estimate for central nuclear amplitudes but not for peripheral ones, see Fig.~\ref{fig:Kaspar_diff}. Also the effect of not including excited proton states in the eikonal have been estimated to be negligible compared to the current experimental precision~\cite{Bethe}. In conclusion, the formulae used for the 13 TeV $\rho$ determination provide more than adequate models for the CNI effects.     

\section{The combination of the $pp$ and $p\bar{p}$ comparison and $\rho$ and $\sigma_{tot}$ measurements}
The significances of the measurements are combined using the Stouffer's method~\cite{Stouffer} in the order of sensitivity, starting from the $pp$ and $p\bar{p}$ comparison, adding the 13 TeV $\rho$ measurement and then finally if needed the $\sigma_{tot}$ measurements using the freedom provided by Stouffer's method to use only a subset of the significances (e.g. $\rho$ and the $pp$ and $p\bar{p}$ comparison) for testing the exclusion of a model. The $\chi^2$ for
the $\sigma_{tot}$ measurements at 2.76, 7, 8 and 13 TeV is computed with respect to the model predictions without odderon exchange~\cite{Compete, Durham, Block_Halzen} including also model uncertainties when specified. Same was done separately for the 13 TeV $\rho$ measurement. Unlike the COMPETE~\cite{Compete} and Block-Halzen~\cite{Block_Halzen} models, the Durham model~\cite{Durham} provides the predicted $d\sigma_{el}/dt$ without odderon exchange contribution. Therefore a direct comparison of the predicted Durham $d\sigma_{el}/dt$ at 1.96 TeV with the D0 $p\bar{p}$ $d\sigma_{el}/dt$ that gives a significance of 4.3$\sigma$ is used for the combined significance instead of the $pp$ and $p\bar{p}$ comparison. The 1.96 TeV $d\sigma_{el}/dt$ of the model is chosen since it is most sensitive to odderon exchange after the model has been tuned to the LHC elastic $pp$ data.

The 13 TeV $\rho$ measurement provides a 4.6 and 3.9$\sigma$ significance for the COMPETE "blue band" (see Fig.~\ref{fig:compete_sigmatot_rho}) and the Block-Halzen models~\cite{Block_Halzen}, respectively. The comparison of $\rho$ and $\sigma_{tot}$ measurements with the predictions of the Durham~\cite{Durham}, the COMPETE "magenta band" and "green band" (see Fig.~\ref{fig:compete_sigmatot_rho}) models give significances of 3.4, 4.0 and 4.6$\sigma$, respectively. Combining them with the significance of the $pp$ and $p\bar{p}$ comparison (or for Durham the one with D0) give combined significances ranging
from 5.2 to 5.7$\sigma$ for odderon exchange for all examined models~\cite{Compete, Durham, Block_Halzen}.

\subsection{Questions and objections raised about the combination}
The Stouffer's method~\cite{Stouffer} combines significances following $z_{\rm comb} = \sum^{k}_{i=1} z_i/\sqrt{k}$, where $z_i$ is the individual significances and $k$ the number of significances to be combined. The method is valid for independent measurements, whose significances obey the normal distribution. This is true for the odderon significances obtained from the $pp$ and $p\bar{p}$ comparison in the dip-bump region and the $pp$ $\rho$ and $\sigma_{tot}$ measurements at very low $|t|$, since they are based on results from completely separate $|t|$ regions and TOTEM data sets. When the 13 TeV  $\rho$ and $\sigma_{tot}$ measurements are both used for the combined significance, values determined from independent TOTEM data sets are used. 

It has also been questioned whether the $pp$ and $p\bar{p}$ comparison and the $\rho$ and $\sigma_{tot}$ measurements can be combined, since the former is a data to data comparison and the latter a data to model comparison. However, since the only way to produce a significant difference between the $pp$ and $p\bar{p}$ $d\sigma_{el}/dt$ at TeV energy scale is through odderon exchange, a model without odderon exchange would produce a $p\bar{p}$ $d\sigma_{el}/dt$ at 1.96 TeV similar to the extrapolated $pp$ $d\sigma_{el}/dt$ if the model still has to describe the $pp$ $d\sigma_{el}/dt$'s measured at LHC. This is illustrated by the Durham model without odderon contribution that fails to describe the D0 $p\bar{p}$ $d\sigma_{el}/dt$ at 1.96 TeV (at a 4.3$\sigma$ significance). Also the failure of the models to describe simultaneously both the $\rho$ and $\sigma_{tot}$ measurements in $pp$ points to a difference in elastic $pp$ and $p\bar{p}$ scattering and therefore to be quantitatively assessing the same thing, the existence of odderon exchange in elastic scattering, as the $pp$ and $p\bar{p}$ comparison. 

\section{Conclusions}
Issues and objections raised regarding the D0-TOTEM comparison of the elastic $d\sigma/dt$ of $pp$ and $p\bar{p}$, the TOTEM $\rho$ and $\sigma_{tot}$ measurements in $pp$ as well as their combination and odderon interpretation have been adequately addressed. Both provide evidence of odderon exchange in elastic scattering and their combination constitute the first experimental observation of the odderon, acknowledged as convincing evidence of the existence of the odderon after a quest of almost 50 years~\cite{Leader}.

\end{document}